\documentclass[aps, prd, 12pt, eqsecnum, tightenlines, notitlepage, nofootinbib, floatfix]{revtex4-1}

\PassOptionsToPackage{unicode}{hyperref}

\usepackage{hyperref, graphicx, slashed, xcolor, amsmath}

\allowdisplaybreaks

\definecolor{red}{rgb}{1,0,0}

\newcommand{\beq}{\begin{equation}}
\newcommand{\eeq}{\end{equation}}
\newcommand{\bea}{\begin{eqnarray}}
\newcommand{\eea}{\end{eqnarray}}
\newcommand{\nn}{\nonumber\\}

\begin{document}

\baselineskip=18pt \pagestyle{plain} \setcounter{page}{1}

%\preprint{
\begin{flushright}
CALT-TH-2018-027
\end{flushright}
%}

\vspace*{1.5cm}

\title{De Sitter Quantum Loops as the origin of Primordial Non-Gaussianities}

\author{Haipeng An$^a$, Mark B. Wise$^b$ and Zipei Zhang$^a$}
\affiliation{$^a$ Department of Physics, Tsinghua University, Beijing 100084, China}
\affiliation{$^b$Walter Burke Institute for Theoretical Physics,
California Institute of Technology, Pasadena, CA 91125}

\begin{abstract}

It was pointed out recently that in some inflationary models quantum loops containing a  scalar of mass $m$ that couples to the inflaton can be the dominant source of  primordial non-Gaussianities. We explore this phenomenon in the simplest such model focusing on the behavior of the primordial curvature fluctuations for small $m/H$. Explicit calculations are done for the three and four point curvature fluctuation correlations. Constraints on the parameters of the model from the CMB limits on primordial non-Gaussianity are discussed. The bi-spectrum in the squeezed limit and the tri-spectrum in the compressed limit are examined. The form of the $n$-point correlations as any partial sum of wave vectors gets small is determined. 
\end{abstract}
%\preprint{CALT-TH-2018-027}
\maketitle

\section{Introduction}

An inflationary era in the early universe is a popular possibility for the solution of the horizon and flatness problems~\cite{SKS}. It also provides an elegant mechanism to generate primordial density perturbations that, in the early universe, have wavelengths well outside the horizon. In the most simple inflationary cosmology where only a single inflaton field with a standard kinetic term plays a role, the density perturbations are almost Gaussian and any non-Gaussianities will be unobservable for the forseable future~\cite{Maldacena:2002vr}.

The inflationary era occurs in the early universe when the energy density is (temporarily) dominated by vacuum energy resulting in a scale factor that expands exponentially with time. This exponential expansion inflates the size of regions that were in causal contact to enormous size solving the horizon problem. It also increases exponentially the physical wavelength of perturbations with fixed comoving wavelength causing the density perturbations that arise from quantum fluctuations and are relevant for large scale structure and the cosmic microwave background radiation (CMB) to have wavelengths that are well outside the horizon when inflation ends. At the end of the inflationary era the universe reheats to a conventional radiation (or matter) dominated universe.

Galaxies are biased objects and as such the power spectrum for fluctuations\footnote{A similar phenomena happens for the galaxy number density bi-spectrum.}  in their number density can be enhanced, at low wave vectors, if there are non-Gaussian primordial curvature correlations, (generated in the inflationary era) that are enhanced as a single wave vector or partial sum of wave vectors go to zero~\cite{AGW,Dalal:2007cu,Baumann:2012bc,Gleyzes:2016tdh}. For the three and four point primordial curvature correlations the dominant enhancements occur respectively in the squeezed and compressed limits. 

These enhancements in the galaxy number density power spectrum (and bi-spectrum~\cite{An:2017rwo}) at very small wave vectors are  sometimes called scale dependent bias and have been studied extensively in quasi single field inflation (QSFI~\cite{Chen:2009zp}) which contains an additional scalar degree of freedom $s$  with mass $m$. They cannot arise from non-linear gravitational evolution~\cite{Goroff:1986ep}.  

In QSFI the primordial  non-Gaussianity results from tree diagrams in this theory. Recently a model was constructed where it is quantum loop diagrams in de Sitter space that give rise to the non-Gaussian correlations that enhance the galaxy number density power spectrum at small wave vectors~\cite{McAneny:2017bbv}. 

In this paper we follow up on this observation considering the simplest model where loop diagrams dominate the non-Gaussianities. This model contains the inflaton field and an additional massive scalar $s$  with mass $m$ that we take to be small compared with the Hubble constant during inflation. A $Z_2$ symmetry in the system forbids the tree-level contribution from $s$ to the correlation functions of the curvature perturbation. We compute the curvature three point correlation (bi-spectrum) and four point correlation (tri-spectrum) for general wave vectors. When all ratio's of wave vectors are not unusually large or small the bi-spectrum and tri-spectrum respectively have the same form as local non-Gaussianity and $\tau_{NL}$ non-Gaussianity\footnote{ A tri-spectrum with $\tau_{NL}$ non-Gaussianity  satisfies Eq.~(21) of~\cite{Ade:2013ydc}  with $g_{NL}=0$.}. In the squeezed and compressed limits the bi-spectrum and tri-spectrum have the familiar enhancements that give rise to an enhanced galaxy power spectrum at low wave vectors. We also discuss the form of the $n$-point curvature fluctuations as a partial sum of wave vectors goes to zero. We  plot  the enhancement of the two-point power spectrum of the galactic halo distributions generated by the quantum loops in this model using a simplified threshold model~\cite{Gunn:1972sv} for the galaxy halo number density. 

The rest of the paper is organized as the following. In Sec.~\ref{sec:model} we discuss the details of the model. In Sec.~\ref{sec:general} we present the general three and four-point correlation functions of the curvature perturbation.  Details of the calculations are relegated to appendices. In Sec.~\ref{sec:observables} we calculate the constraints on the parameters of the model from CMB limits on  the primordial bi-spectrum and tri-spectrum~\cite{Ade:2013ydc, Ade:2015ava}. In contrast to most models the constraint from the limit on the tri-spectrum is stronger. We conclude in Sec.~\ref{sec:conclusion}.

\section{The Model}
\label{sec:model}

The model contains two scalar fields, the inflaton $\phi$ and another scalar field $s$ of mass $m$. We assume the inflation and $s$ fields have a $\phi \rightarrow -\phi$ symmetry and $s \rightarrow -s$ symmetry. We also assume a shift symmetry for $\phi$ that is only broken by the inflation potential.  Then the lowest dimension term that  contains interactions of $\phi$ with $s$ is \begin{equation}
{\cal L}_{\rm int}={ 1\over 4 \Lambda^2 }g^{\mu \nu} \left( \partial_{\mu} \phi \partial_{\nu} \phi \right) s^2
\end{equation}
During inflation $\phi$ has a background value $\phi_0(t)$ that depends on time. Its magnitude relative to $H^2$ is fixed by the CMB temperature fluctuations to be,  ${\dot \phi_0}/H ^2\simeq  3.5\times10^4$.

This model has a fine tuning. The physical $s$ mass $m$ is a sum of two terms one the mass parameter $\mu$ from the $s$  potential and the other from the interaction above. Explicitly,
\begin{equation}
\label{cancel}
m^2= \mu^2-{{\dot \phi}_0 ^2\over 4 \Lambda^2}
\end{equation}
To get the enhancements we mentioned in the introduction we need $m/H < 1$ where $H$ is the Hubble constant during inflation. But to get observable non-Gaussianities we need
the second term in the equation above to be much greater than $H^2$, hence the tuning.  Finally we assume that there exists an inflaton potential that gives an exceptable region in $r, n_s$ space and is flat enough that $m$ can be approximated as a constant for the calculation of the non-Gaussianities generated during inflation.

We proceed along the lines of the effective field theory of inflation working in the gauge where $\phi(x) = \phi_0(t)$~\cite{Cheung:2007st}. A Goldstone mode $\pi$, due to the breaking of the time translation symmetry, is introduced to describe the curvature perturbation as follows,
\beq
g^{\mu\nu} \partial_\nu\phi \partial_\nu\phi \rightarrow g^{00} \dot\phi_0^2 \rightarrow \dot\phi_0^2g^{\mu\nu} \partial_\nu(t+\pi/\dot\phi_0) \partial_\nu(t+\pi/\dot\phi_0)  \ ,
\eeq
where the evolution of $\dot\phi_0$ is neglected and the relation between $\pi$ and the curvature perturbation $\zeta$ in this model can be written as
\beq
\zeta = -\frac{H}{\dot\phi_0} \pi \ .
\eeq
A three-point and a four-point interaction between $\pi$ and $s$ are induced with 
\beq
{\cal L}_{\rm int} = - \frac{1}{2 H^3 \tau^3} \frac{\dot\phi_0}{\Lambda^2} \frac{\partial\pi}{\partial\tau} s^2 + \frac{1}{ 4H^2\tau^2 \Lambda^2} \eta^{\mu\nu} \partial_\nu\pi \partial_\nu \pi s^2 \ ,
\eeq
where $\tau$ is the conformal time defined as $\tau = - e^{-H t}/H$ and $\eta^{\mu\nu}$ = diag(1,-1,-1,-1). Here we include the $\sqrt{-g}$ factor into the Lagrangian. 

The mode expansion of $\pi$ and $s$ are defined as
\bea
\pi(\vec x,\tau) &=& \int\frac{d^3k}{(2\pi)^3} \left[ \pi_k(\tau) e^{i \vec k \cdot \vec x} a_{\vec k} + \pi^{*}_k(\tau) e^{-i \vec k \cdot \vec x} a_{\vec k}^\dagger \right] \nn
s(\vec x,\tau) &=& \int\frac{d^3k}{(2\pi)^3} \left[ s_k(\tau) e^{i \vec k \cdot \vec x} b_{\vec k} + s^{*}_k(\tau) e^{-i \vec k \cdot \vec x} b_{\vec k}^\dagger \right] 
\eea
\beq
[a_{\vec k}, a^\dagger_{\vec k'}] = [b_{\vec k}, b^\dagger_{\vec k'}] = (2\pi)^3\delta(\vec k - \vec k') \ .
\eeq
Assuming the Bunch-Davies vacuum, the mode functions can be written as
\begin{equation}
\pi_k = \frac{H}{\sqrt{2k}} \left( i\tau + \frac{1}{k} \right) e^{-ik\tau} 
\end{equation}
and
\begin{equation}
s_k =  {H (-k\tau)^{3/2} \pi^{1/2} \over  2 k^{3/2}}  H^{(2)}_\alpha (k\tau),
\end{equation} 
where 
\beq
\alpha = \left(\frac{9}{4} - \frac{m^2}{H^2}\right)^{1/2} \ ,
\eeq
and $H_\alpha^{(2)}$ is the second Hankel function with index $\alpha$. 
Throughout this paper we focus on the region where $m^2/H^2$ is small compared with unity. For  $|k \tau| \ll 1$,
\begin{equation}
s_k(\eta)=  {H \pi^{1/2} \over 2k^{3/2}} (-k\tau)^{\nu} a_0
\end{equation}
where,
\begin{equation}
 \nu =3/2- \alpha  \simeq m^2/(3H^2)
 \end{equation} 
 and
\begin{equation}
a_0= i{2^{3/2-\nu} \Gamma(3/2-\nu) \over \pi} \simeq {2^{1/2}i \over \pi^{1/2}}.
\end{equation}
We are interested in the multi-point correlation functions of the curvature perturbation far outside the horizon $(\tau \simeq 0)$, so we need to evaluate $\langle \zeta(\vec x_1,0) \cdots \zeta(\vec x_n,0)\rangle$. This can be calculated using~\cite{Weinberg:2005vy}
\beq\label{eq:master}
\langle {\cal O}(0) \rangle = \sum_N i^N \int^0_{-\infty} d\tau_N \int_{-\infty}^{\tau_N} d\tau_{N-1} \cdots \int_{-\infty}^{\tau_{2}} d\tau_1 \langle [H^I_{\rm int}(\tau_1),[H^I_{\rm int}(\tau_1),\cdots H^I_{\rm int}(\tau_N),{\cal O}^I(0)] \cdots ]\rangle\ ,
\eeq 
where the $I$ in the superscript stands for interaction picture.

\section{General form of the three and four-point correlation functions of $\zeta$}
\label{sec:general}

\subsection{Three-point correlation function}

\begin{figure}
\centering
\includegraphics[height=2.5in]{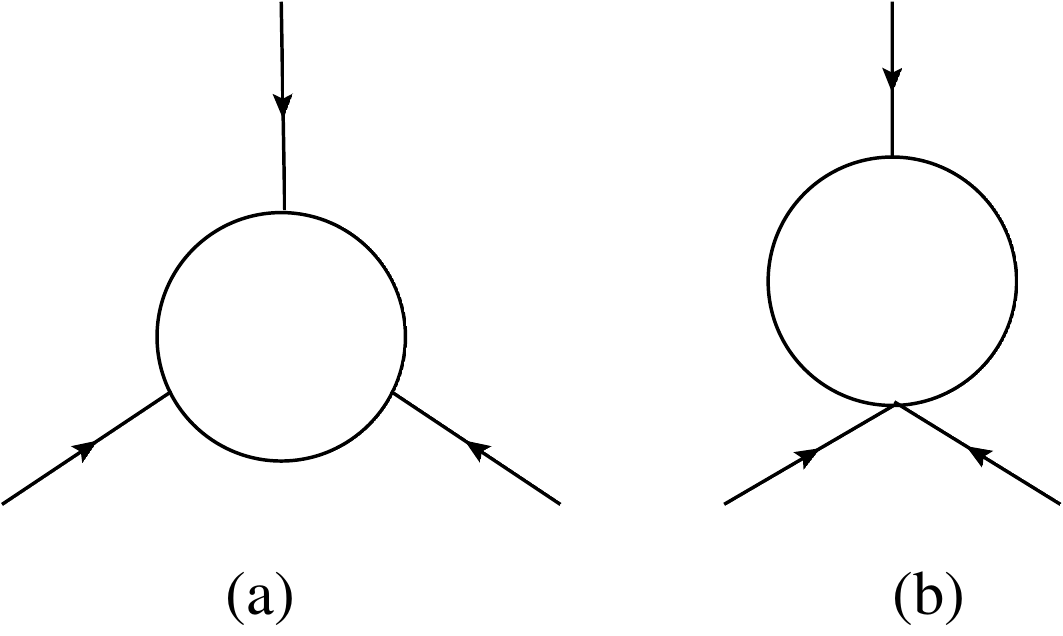}
\caption{}\label{fig:3pt}
\end{figure}

There are  two diagrams generating one-loop contributions to the three-point function of $\pi$, which are shown in Fig.~\ref{fig:3pt}. We will show that with the saturation of the current limits of $f_{\rm NL}$ and $\tau_{\rm NL}$ the contribution from Fig.~\ref{fig:3pt}(a) is much larger than from the Fig.~\ref{fig:3pt}(b). The calculation of the resulting bi-spectrum is performed in Appendix A and there it was found that for $m/H \ll 1$
 
\bea\label{eq:B3general2}
{\cal B}_\zeta(\vec k_1 , \vec k_2 , \vec k_3) &=& \frac{\pi |a_0|^6 H^6}{128\Lambda^6} \frac{1}{(2\nu)^4} \nn
&&\!\!\!\!\!\!\!\!\!\!\!\!\!\!\!\!\!\!\!\!\!\!\!\!\!\!\times\left\{ \frac{\max(k_1,k_3)^{-2\nu} k_{\max}^{-2\nu}}{k_1^{3-2\nu} k_3^{3-2\nu}} + \frac{\max(k_2,k_3)^{-2\nu} k_{\max}^{-2\nu}}{k_2^{3-2\nu} k_3^{3-2\nu}} + \frac{\max(k_1,k_2)^{-2\nu} k_{\max}^{-2\nu}}{k_1^{3-2\nu} k_2^{3-2\nu}}\right\}
\eea
This is  the general leading order expression for the bi-spectrum of $\zeta$ .

A factor $(k_i/k_j)^{\nu} \simeq  1+\nu { \rm ln} (k_i/k_j) + ... .$. So  such factors can be set to unity when all the ratios of the $k$'s are  not very small or large. Then

\begin{equation}
\label{eq:B3typical}
{\cal B}_\zeta(\vec k_1 , \vec k_2 , \vec k_3) \simeq  \frac{\pi |a_0|^6 H^6}{128\Lambda^6} \frac{1}{(2\nu)^4}
\times\left\{ \frac{1}{k_1^{3} k_3^{3}} + \frac{1}{k_2^{3} k_3^{3}} + \frac{1}{k_1^{3} k_2^{3}}\right\}
\end{equation}
This is the form that the bi-spectrum has in local non-Gaussianity.

Next we consider the squeezed region first where $ k \equiv k_1 \simeq k_2 \gg k_3 \equiv q$, it is easy to see that in this limit the bi-spectrum goes to

\bea\label{eq:B3sq}
{\cal B}_\zeta(\vec k_1 , \vec k_2 , \vec k_3)   &\simeq&  \frac{\pi H^6 |a_0|^6}{64\Lambda^6} \frac{1}{(2\nu)^4} \frac{1}{k^{3+2\nu} q^{3- 2\nu}} \ .
\eea
This differs from what one gets from local non-Gaussianity by a factor of $(q/k)^{2 \nu}$.

\subsection{Four-point correlation function}

\begin{figure}
\centering
\includegraphics[height=1.7in]{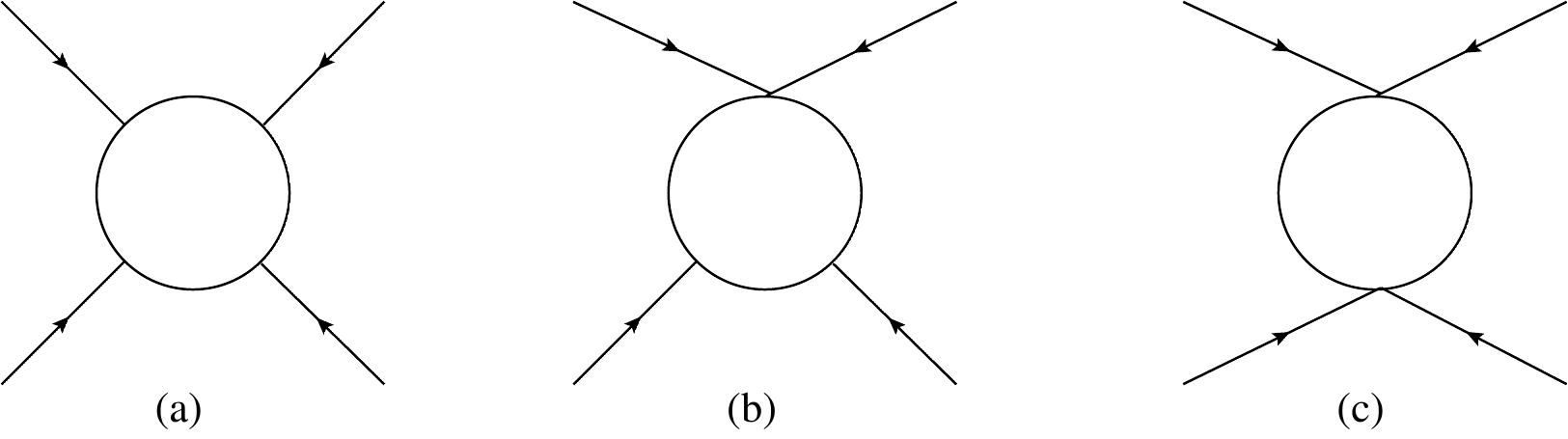}
\caption{}\label{fig:4pt}
\end{figure}

There are  three Feynman diagrams contributing to the one-loop four-point function of $\zeta$, which are shown in Fig.~\ref{fig:4pt}.  We will show that with the saturation of the current limits of $f_{\rm NL}$ and $\tau_{\rm NL}$ the contribution from  Fig.~\ref{fig:4pt}(a) dominates. In Appendix  B the contribution from Fig.~\ref{fig:4pt}(a) was found to be,
\bea
\label{eq:3_4}
&&{\cal B}_{\zeta}(\vec k_1, \vec k_2, \vec k_3, \vec k_4) \nn
&=& \frac{\pi^2 H^8 |a_0|^8}{1024 \Lambda^8} \left(\frac{1}{2\nu}\right)^5\nn
&&\times 
\frac{\min[\max(k_1,k_2,|\vec k_1 + \vec k_2|),\max(k_3, k_4,|\vec k_3 + \vec k_4|)]^{-2\nu} \max(k_1,k_2,k_3,k_4,|\vec k_1 + \vec k_2|)^{-2\nu}}{k_1^3 k_4^3 |\vec k_1 + \vec k_2|^{3-2\nu}  \min(k_1,k_4, |\vec k_1 + \vec k_2|)^{-2\nu}} \nn 
&&+ {\rm per}(k_1, k_2, k_3, k_4)  \ .
\eea

Provided non of the ratio's of the $k$'s or partial sums of the $k$'s is anomalously  large this has the same form as  $\tau_{NL}$ ($g_{NL}=0$) non-Gaussianity,

\begin{equation}
\label{eq:3_5}
{\cal B}_{\zeta}(\vec k_1, \vec k_2, \vec k_3, \vec k_4) \simeq \frac{\pi^2 H^8 |a_0|^8}{1024 \Lambda^8} \left(\frac{1}{2\nu}\right)^5
\left( \frac{1}{k_1^3 k_4^3 |\vec k_1 + \vec k_2|^{3} } 
 + {\rm per}(k_1, k_2, k_3, k_4) \right) \ .
\end{equation}

In the compressed region $k_1 \simeq k_2 \gg |\vec k_1 + \vec k_2| \equiv q$ and $k_3 \simeq k_4 \gg q$, it simplifies into 
\beq\label{eq:B4compressed}
{\cal B}(\vec k_1, \vec k_2, \vec k_3, \vec k_4) \simeq \frac{\pi^2 H^8 |a_0|^8}{128 \Lambda^8} \left(\frac{1}{2\nu}\right)^5 \frac{1}{k_1^{3 + 2\nu}} \frac{1}{k_3^{3 + 2\nu}} \frac{1}{q^{3 - 4\nu}} \ .
\eeq

The scaling of the four point curvature fluctuation in the compressed limit ($\sim q^{-3+4\nu}$) can be read out from the scaling dimension of the operator $s^2$ since this determines the form of its two point correlation evaluated on the boundary $\tau=0$ of de Sitter space.~\cite{Arkani-Hamed:2015bza}.

\section{CMB Constraints}
\label{sec:observables}

In the previous section it was noted that for  $\nu \ll 1$, the bi-spectrum of this model is the same as the local non-Gaussianity model (for typical wave vectors). Therefore, we can use the observed limit of the $f_{\rm NL}^{\rm local}$ to estimate the constraint on this model the CMB observation of the non-Gaussianity. In the limit where the bi-spectrum reduces to local non-Gaussianity,
\bea
f_{\rm NL}^{\rm local} &=& {5 \over 6} \left(\frac{{\cal B}_{\zeta}(k_1, k_2 ,k_3)}{P_{\zeta}(k_1) P_{\zeta}(k_2)+P_{\zeta}(k_1) P_{\zeta}(k_3)+P_{\zeta}(k_2) P_{\zeta}(k_3)}\right) \\ \nonumber
&\simeq& \frac{5}{6} \frac{1}{2^9\pi^3} \left(\frac{H}{\Lambda}\right)^6 |a_0|^6 (\Delta_\zeta^2)^{-2}  \left(\frac{1}{2\nu}\right)^4 \ ,
\eea
where 
\bea\label{eq:Delta2}
\Delta^2_\zeta \equiv \frac{1}{(2\pi)^2} \frac{H^4}{\dot\phi_0^2} \simeq 2.14\times 10^{-9} \ .
\eea
The $2\sigma$ constraint~\cite{Ade:2015ava}  on $f_{\rm NL}^{\rm local}$ is about $|f_{\rm NL}^{\rm local}| < 10$. Therefore, we have 
\beq
\frac{H}{\Lambda} \left(\frac{1}{2\nu}\right)^{2/3} < 0.012. \ 
\eeq

Similarly for the tri-spectrum $\tau_{\rm NL}$ can be estimated as
\bea
\tau_{\rm NL} \simeq \frac{H^8 |a_0|^8}{4096 \pi^4 \Lambda^8} \left(\frac{1}{2\nu}\right)^5 (\Delta_\zeta^2)^{-3}  \ .
\eea
The constraint on $\tau_{\rm NL}$ from the Planck observation of the CMB spectrum~\cite{Ade:2013ydc} is
\bea
\tau_{\rm NL} < 2800 \ ,
\eea
which implies that
\bea
\label{limit}
\frac{H}{\Lambda} \left(\frac{1}{2\nu}\right)^{5/8}  < 0.0095 \ .
\eea

We can use these constraints to estimate the importance of diagram 2b in comparison with 2a for the four point correlation. It is straightforward to see that,
\begin{equation}
{{\rm diagram 2a} \over {\rm diagram 2b}} \sim {{\dot \phi_0}^2 \over H^4} {H^2 \over \Lambda^2}{1 \over \nu}.
\end{equation}
Treating Eq.~(\ref{limit}) as an equality this becomes
\begin{equation}
{{\rm diagram 2a} \over {\rm diagram 2b}} \sim 10^{-4}{ {\dot \phi_0}^2 \over H^4}\nu^{1/4}.
\end{equation}
which is much larger than unity unless $\nu$ is exceptionally small\footnote{Although we have only calculated our results to leading order in $1/\nu$, $\nu \sim 0.1$ should be small enough for these results to be a reasonable approximation.}. Similar conclusions hold for diagram 2c and for other correlation functions.

\begin{figure}
\centering
\includegraphics[height=2.7in]{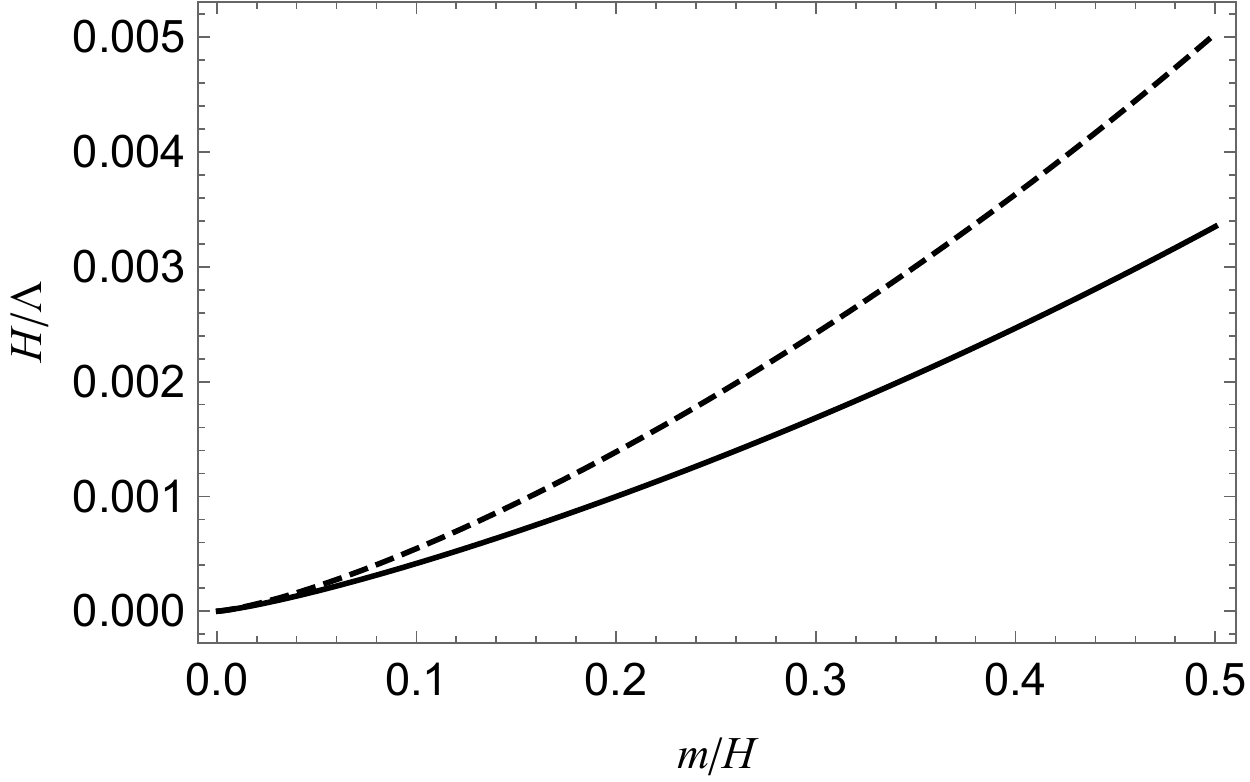}
\caption{The dashed and solid curves show the upper limits on $H/\Lambda$ as a function of $m/H$ from $f_{\rm NL}$ and $\tau_{\rm NL}$.  }\label{fig:constraints}
\end{figure}

The constraints from $f_{\rm NL}^{\rm local}$ and $\tau_{\rm NL}$ are shown in Fig.~\ref{fig:constraints}, where one can see that the constraint from $\tau_{\rm NL}$ is always stronger. This result is in contrast to many other models, where the constraint from $f_{\rm NL}$ is stronger. The main reason is the following. In the models where there are tree-level contributions to non-Gaussianity, $f_{\rm NL}$ is usually proportional to a single power of the small coupling of the new interaction and $\tau_{\rm NL}$ proportional to the square of it. Whereas in this model $f_{\rm NL}$ is proportional to the third power of the coupling and $\tau_{\rm NL}$ proportional to the fourth power of it. Therefore, relatively in this model $\tau_{\rm NL}$ is more important. 

Since there is some small time dependence in ${\dot \phi}_0$,  Eq.~(\ref{cancel}) implies that the $s$ mass depends on time

\begin{equation}
{d{ \rm ln}m^2 \over Hdt}  \simeq {1 \over 24 \pi^2 \Delta_{\zeta} ^2}   \left({H^2 \over \Lambda^2} \right){\eta \over \nu}
\end{equation}

The slow role parameter $\eta$ must be small enough that this time dependence can be neglected in our computations.

In Appendix~\ref{app:d} we calculate the one-loop correction to $\Delta^2_\zeta$ and find that
\bea
\delta\Delta^2_\zeta = \frac{H^4 |a_0|^4}{512 \pi^2 \nu^3 \Lambda^4}  \ .
\eea
The constraints from $\tau_{\rm NL}$ and $f_{\rm NL}$ are
\bea
\delta\Delta^2_\zeta < \min \left[6\times 10^{-12} \times \left(\frac{1}{2\nu}\right)^{1/2} , 1.3\times 10^{-11} \times  \left(\frac{1}{2\nu}\right)^{1/3} \right]\ ,
\eea
which is much smaller than the observed value given in Eq.~(\ref{eq:Delta2}) unless $\nu$ is exceptionally small$^{3}$.
 
\section{Concluding Remarks}
\label{sec:conclusion}

We have examined the primordial curvature perturbations in an inflationary cosmology where the inflaton  $\phi$ couples to an additional scalar field $s$ with mass $m\ll H$ through the non-renormalizable interaction $ g^{\mu \nu}\partial_{\mu} \phi \partial_{\nu} \phi/4\Lambda^2$. In this model primordial non-Gaussianities arise from quantum loop diagrams (with a virtual s in the loop) in de Sitter space. The primordial curvature fluctuation bi-spectrum and tri-spectrum in this model were computed.  Typically they respectively have the form of local non-Gaussianity  and $\tau_{NL}$ non-Gaussianity, although in squeezed and compressed limits there are deviations from that form.

It is not difficult given the work of this paper to deduce the form of the $n>4$ curvature perturbations.. In the situation where all the wave vectors and their partial sums of wave vectors are of order $k$,
\begin{equation}
B_{\zeta}(\vec{ k}_1, \ldots , \vec{ k}_n   ) \sim \left({H \over \Lambda}\right)^{2 n} \left( {1 \over 2 \nu}\right)^{n+1}{1 \over k^{3n-3}},
\end{equation}
and in the limit where a single partial sum of wave vectors $|\vec{ k}_1 + \ldots  +\vec{ k}_j|=q \ll k$ it becomes,
\begin{equation}
B_{\zeta}(\vec{ k}_1, \ldots , \vec{ k}_n   ) \sim \left({H \over \Lambda}\right)^{2 n} \left( {1 \over 2 \nu}\right)^{n+1}{1 \over k^{3n-6+4 \nu}}{1 \over q^{3-4 \nu}}.
\end{equation}

\begin{figure}
\centering
\includegraphics[height=2.7in]{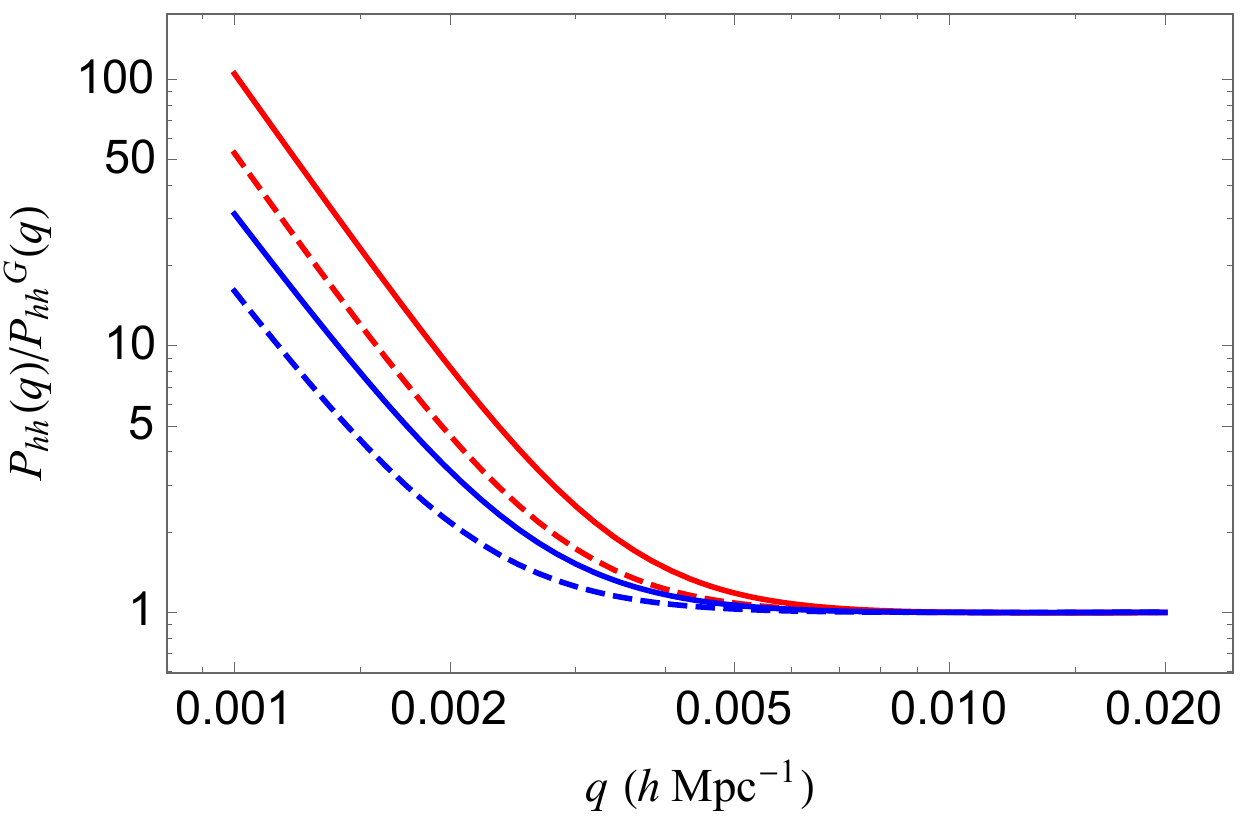}
\caption{The red solid and dashed curves are for $\nu = 0.05$ with $\tau_{\rm NL}= 2800$ and 1400 respectively, and the blue solid and dashed curves for $\nu=0.1$ with the same choices of $\tau_{\rm NL}$. }\label{fig:numerical}
\end{figure}

In this model due to the IR behavior of the compressed tri-spectrum~(\ref{eq:B4compressed}), the long distance behavior of the power spectrum for fluctuations in the galaxy number density is enhanced by a factor of $q ^{4 - 4\nu}$ compared to the Harrison-Zel'dovich spectrum (which goes, apart from a small tilt, as $q$). Following the same procedure as described in Refs.~\cite{An:2017rwo,McAneny:2017bbv} we estimate this power spectrum and its ratio to the leading Harrison-Zel'dovich contribution is shown in Fig.~\ref{fig:numerical} using the curvature bi-spectrum and tri-spectrum calculated in this paper. One can see that if the current constraint from $\tau_{\rm NL}$ is saturated the power spectrum of the galactic halo distribution can differ significantly from the Harrison-Zel'dovich spectrum at $q\sim h/(250~({\rm Mpc}))$. Those  deviations from what Gaussian primordial fluctuations would give can become very large on scales around $500 h^{-1}$ Mpc.

\acknowledgments

HA is supported by the Recruitment Program for Young Professionals of the 1000 Talented Plan and the Tsinghua University Initiative Scientific Research Program. MBW is supported by the DOE Grant DE-SC0011632 and the Walter Burke Institute for Theoretical Physics. 

\appendix

\section{The bi-spectrum}

The contribution from Fig.~\ref{fig:3pt}(a) can be written as
\bea
&&{\cal A}_3^{(a)} (\vec x_1, \vec x_2, \vec x_3) \equiv \langle \zeta(\vec x_1, 0) \zeta(\vec x_1,0) \zeta(\vec x_3,0)\rangle^{(a)}  \nn
&=& \frac{i}{8\Lambda^6 H^6} \int_{-\infty}^0 \frac{d \tau_3}{\tau_3^3} \int_{-\infty}^{\tau_3} \frac{d \tau_2}{\tau_2^3} \int_{-\infty}^{\tau_2} \frac{d \tau_1}{\tau_1^3} \int d^3 y_1d^3 y_2d^3 y_3 \nn
&&\times \langle \left[\pi'(\tau_1,\vec y_1)s^2(\tau_1,\vec y_1),\left[\pi'(\tau_2,\vec y_2)s^2(\tau_2,\vec y_2),\left[\pi'(\tau_3,\vec y_3)s^2(\tau_3,\vec y_3), \pi(0,\vec x_1)\pi(0,\vec x_2)\pi(0,\vec x_3)\right]\right]\right] \rangle \nn
\eea
Here $'$ means derivative over the conformal time $\tau$. The calculation can be done step by step from inside to the outside of the commutation relations in the following way. 
\bea
&&\left[\pi'(\tau_3,\vec y_3)s^2(\tau_3,\vec y_3), \pi(0,\vec x_1)\pi(0,\vec x_2)\pi(0,\vec x_3)\right] \nn
&=& \left[ \pi'(\tau_3,\vec y_3),\pi(0,\vec x_1)\right] s^2(\tau_3,\vec y_3) \pi(0,\vec x_2)\pi(0,\vec x_3) + (x_1 \leftrightarrow x_2) + (x_1 \leftrightarrow x_3)
\eea
Note that $\left[ \pi'(\tau_3,\vec y_3),\pi(0,\vec x_1)\right]$ is a number we have
\beq
\left[ \pi'(\tau_3,\vec y_3),\pi(0,\vec x_1)\right] = \langle \left[ \pi'(\tau_3,\vec y_3),\pi(0,\vec x_1)\right] \rangle = 2i{\rm Im}\langle \pi'(\tau_3,\vec y_3)\pi(0,\vec x_1) \rangle \ .
\eeq
Therefore we have
\bea
&&\left[\pi'(\tau_2,\vec y_2)s^2(\tau_2,\vec y_2),\left[\pi'(\tau_3,\vec y_3)s^2(\tau_3,\vec y_3), \pi(0,\vec x_1)\pi(0,\vec x_2)\pi(0,\vec x_3)\right]\right]\nn
&=& 2i{\rm Im}\langle \pi'(\tau_3,\vec y_3)\pi(0,\vec x_1) \rangle \left[\pi'(\tau_2,\vec y_2)s^2(\tau_2,\vec y_2), s^2(\tau_3,\vec y_3) \pi(0,\vec x_2)\pi(0,\vec x_3)\right] \ .
\eea
From the structure of Fig.~\ref{fig:3pt}(a) we find that one $s$ at $(\tau_2,\vec y_2)$ should contract with one $s$ at $(\tau_3,\vec y_3)$ with the other $s$'s contract with the $s^2$ at $(\tau_1,\vec y_1)$. Therefore, we have
\bea
&&\left[\pi'(\tau_2,\vec y_2)s^2(\tau_2,\vec y_2), s^2(\tau_3,\vec y_3) \pi(0,\vec x_2)\pi(0,\vec x_3)\right]_{\rm Fig.~1(a)} \nn
&=& 4 \langle\left[\pi'(\tau_2,\vec y_2)s(\tau_2,\vec y_2), s(\tau_3,\vec y_3) \pi(0,\vec x_2)\right] \rangle s(\tau_2,\vec y_2) s(\tau_3,\vec y_3) \pi(0,\vec x_3) + (x_2 \leftrightarrow x_3) \nn
&=& 8i {\rm Im} \left[ \langle \pi'(\tau_2,\vec y_2) \pi(0,\vec x_2)\rangle \langle s(\tau_2,\vec y_2) s(\tau_3, \vec y_3) \rangle \right] s(\tau_2,\vec y_2) s(\tau_3,\vec y_3) \pi(0,\vec x_3)+ (x_2 \leftrightarrow x_3)
\eea
Therefore, we have for the contribution from Fig.~\ref{fig:3pt}(a)
\bea
&&\langle \left[\pi'(\tau_1,\vec y_1)s^2(\tau_1,\vec y_1),\left[\pi'(\tau_2,\vec y_2)s^2(\tau_2,\vec y_2),\left[\pi'(\tau_3,\vec y_3)s^2(\tau_3,\vec y_3), \pi(0,\vec x_1)\pi(0,\vec x_2)\pi(0,\vec x_3)\right]\right]\right] \rangle \nn
&=& -16  {\rm Im}\langle \pi'(\tau_3,\vec y_3)\pi(0,\vec x_1) \rangle {\rm Im} \left[ \langle \pi'(\tau_2,\vec y_2) \pi(0,\vec x_2)\rangle \langle s(\tau_2,\vec y_2) s(\tau_3, \vec y_3) \rangle \right] \nn 
&&\times \langle \left[\pi'(\tau_1,\vec y_1)s^2(\tau_1,\vec y_1), s(\tau_2,\vec y_2) s(\tau_3,\vec y_3) \pi(0,\vec x_3) \right] \rangle + {\rm per(x_1,x_2,x_3)} \nn
&=& -64i {\rm Im}\langle \pi'(\tau_3,\vec y_3)\pi(0,\vec x_1) \rangle {\rm Im} \left[ \langle \pi'(\tau_2,\vec y_2) \pi(0,\vec x_2)\rangle \langle s(\tau_2,\vec y_2) s(\tau_3, \vec y_3) \rangle \right] \nn
&&\times {\rm Im}\left[ \langle \pi'(\tau_1,\vec y_1) \pi(0,\vec x_3)\rangle \langle s(\tau_1,\vec y_1)s(\tau_2,\vec y_2)\rangle \langle s(\tau_1,\vec y_1)s(\tau_3,\vec y_3)\rangle \right] + {\rm per(x_1,x_2,x_3)}
\eea
Now, take the second factor as an example,
\bea
&&{\rm Im} \left[ \langle \pi'(\tau_2,\vec y_2) \pi(0,\vec x_2)\rangle \langle s(\tau_2,\vec y_2) s(\tau_3, \vec y_3) \rangle \right] \nn
&=& \int \frac{d^3p}{(2\pi)^3} \frac{d^3q}{(2\pi)^3} {\rm Im} \left[ \pi'_p (\tau_2) \pi_p^*(0) e^{i \vec p \cdot(\vec y_2 - \vec x_2)} s_q(\tau_2) s^*_q(\tau_3) e^{i \vec q \cdot (\vec y_2 - \vec y_3)}\right] \ .
\eea
Since the mode functions $\pi_p$ and $s_q$ are even functions of $\vec p$ and $\vec q$ we can move the exponentials outside. Therefore, we have
\bea
&&{\rm Im} \left[ \langle \pi'(\tau_2,\vec y_2) \pi(0,\vec x_2)\rangle \langle s(\tau_2,\vec y_2) s(\tau_3, \vec y_3) \rangle \right] \nn
&=& \int \frac{d^3p}{(2\pi)^3} \frac{d^3q}{(2\pi)^3} e^{i \vec p \cdot(\vec y_2 - \vec x_2) + i \vec q \cdot (\vec y_2 - \vec y_3) } {\rm Im} \left[ \pi'_p (\tau_2) \pi_p^*(0)  s_q(\tau_2) s^*_q(\tau_3) \right] \ .
\eea
Therefore, we have
\bea\label{eq:A3}
&&{\cal A}_3^{(a)} (\vec x_1, \vec x_2, \vec x_3) \nn
&=& \frac{8}{\Lambda^6 H^6} \int_{-\infty}^0 \frac{d \tau_3}{\tau_3^3} \int_{-\infty}^{\tau_3} \frac{d \tau_2}{\tau_2^3} \int_{-\infty}^{\tau_2} \frac{d \tau_1}{\tau_1^3} \int \frac{d^3k_1}{(2\pi)^3}\frac{d^3k_2}{(2\pi)^3} \frac{d^3k_3}{(2\pi)^3} (2\pi)^3 \delta^3(\vec k_1 + \vec k_2 + \vec k_3) e^{i(\vec k_1 \cdot x_1 + \vec k_2 \cdot x_2 + \vec k_3 \cdot x_3)}\nn
&&\int\frac{d^3p}{(2\pi)^3}\times {\rm Im}[\pi'_{k_1}(\tau_3) \pi^*_{k_1}(0)] {\rm Im}\left[ \pi'_{k_2}(\tau_2) \pi^*_{k_2}(0) s_{|\vec p + \vec k_1|}(\tau_2) s^*_{|\vec p + \vec k_1|} (\tau_3) \right] \nn
&&\times {\rm Im}\left[ \pi'_{k_3}(\tau_1) \pi^*_{k_3}(0) s_{|\vec p - \vec k_3|}(\tau_1) s^*_{|\vec p - \vec k_3|}(\tau_2)  s_p(\tau_1) s^*_p(\tau_3)   \right] + {\rm per}(x_1, x_2, x_3) \ .
\eea
The corresponding part of the bi-spectrum of $\zeta$, ${\cal B}^{(a)}_\zeta(\vec k_1, \vec k_2, \vec k_3)$ is defined as 
\beq
{\cal A}_3^{(a)} (\vec x_1, \vec x_2, \vec x_3) = \int\frac{d^3k_1}{(2\pi)^3}\frac{d^3k_2}{(2\pi)^3}\frac{d^3k_3}{(2\pi)^3} e^{i(\vec k_1 \cdot \vec x_1 + \vec k_2 \cdot \vec x_2 + \vec k_3 \cdot \vec x_3)} (2\pi)^3\delta^3(\vec k_1 + \vec k_2 + \vec k_3) {\cal B}^{(a)}_\zeta(\vec k_1 , \vec k_2 , \vec k_3) \ ,
\eeq
which is easy to be read out from Eq.~(\ref{eq:A3}) that
\bea\label{eq:B3}
&&{\cal B}^{(a)}_\zeta(\vec k_1 , \vec k_2 , \vec k_3) \nn
&=& \frac{8}{\Lambda^6 H^6} \int_{-\infty}^0 \frac{d \tau_3}{\tau_3^3} \int_{-\infty}^{\tau_3} \frac{d \tau_2}{\tau_2^3} \int_{-\infty}^{\tau_2} \frac{d \tau_1}{\tau_1^3} \int\frac{d^3p}{(2\pi)^3}   {\rm Im}[\pi'_{k_1}(\tau_3) \pi^*_{k_1}(0)]  \nn
&&\times   {\rm Im}\left[ \pi'_{k_2}(\tau_2) \pi^*_{k_2}(0) s_{|\vec p + \vec k_1|}(\tau_2) s^*_{|\vec p + \vec k_1|} (\tau_3) \right] {\rm Im}\left[ \pi'_{k_3}(\tau_1) \pi^*_{k_3}(0) s_{|\vec p - \vec k_3|}(\tau_1) s^*_{|\vec p - \vec k_3|}(\tau_2)  s_p(\tau_1) s^*_p(\tau_3)   \right] \nn
&&+ {\rm per} (k_1, k_2, k_3) \ .
\eea

In the limit of massless $s$ one can show that all the $\tau_i$ integrals experience IR divergences. This tells us that in the region where $m/H \ll 1$ the IR contribution dominates the $\tau_i$ integrals. Therefore, we can capture the main contribution by Laurent expanding the integrand and calculating the contributions from the leading term. 

The Hankel function $H_\alpha^{(2)} (k\tau)$  in general can be expanded into two series that
\beq\label{eq:H2}
H_\alpha^{(2)} (k\tau) = \sum_{n=0}^{\infty} a_n (- k\tau)^{-\alpha+n} + \sum_{n=0}^{\infty} b_n (- k\tau)^{2\alpha + n} \ .
\eeq
The important property is that the coefficients in each series share the same phase. Then after some straightforward calculation the leading contribution of the bi-spectrum of $\zeta$ can be written as
\bea
{\cal B}^{(a)}_\zeta(\vec k_1 , \vec k_2 , \vec k_3) &\simeq&  \frac{\pi^3 H^6 |a_0|^6}{64\Lambda^6} \int\frac{d^3p}{(2\pi)^3} \int_{-\Lambda_3}^0 \frac{d \tau_3}{(-\tau_3)^{1 - 2\nu}} \int_{-\Lambda_2}^{\tau_3} \frac{d \tau_2}{(-\tau_2)^{1 - 2\nu}} \int_{-\Lambda_1}^{\tau_2} \frac{d \tau_1}{(-\tau_1)^{1 - 2\nu}} \nn
&&\times  \frac{1}{|\vec p + \vec k_1|^{3 - 2 \nu}|\vec p - \vec k_3|^{3- 2 \nu} p^{3 - 2\nu}} + {\rm per}(k_1, k_2, k_3)\ .
\eea 
The UV cut-off of each $\tau_i$ integral is determined by the point where the integrant start to oscillate that
\beq
\Lambda_1 = \min(k_3^{-1}, |\vec p - \vec k_3|^{-1}, p^{-1}),\;\Lambda_2 = \min(k_2^{-1}, |\vec p + \vec k_1|^{-1}, |\vec p - \vec k_3|^{-1}), \;\Lambda_3 = \min(k_1^{-1}, |\vec p + \vec k_1|^{-1}, p^{-1}).
\eeq
Now we can do the $\tau_i$ integrals explicitly and we have
\bea\label{eq:general3}
{\cal B}^{(a)}_\zeta(\vec k_1 , \vec k_2 , \vec k_3) = \frac{\pi^3 H^6 |a_0|^6}{64\Lambda^6} \frac{1}{(2\nu)^3}\int\frac{d^3p}{(2\pi)^3}   \frac{{\cal I}_3(\Lambda_1, \Lambda_2, \Lambda_3)}{|\vec p + \vec k_1|^{3 - 2 \nu}|\vec p - \vec k_3|^{3- 2 \nu}p^{3 - 2\nu}} + {\rm per}(k_1, k_2, k_3)\ , \nn
\eea
where
\bea
{\cal I}_3(\Lambda_1, \Lambda_2, \Lambda_3) = (\Lambda_1 \Lambda_{12} \Lambda_{123})^{2\nu} - \frac{1}{2} \Lambda_{12}^{4\nu} \Lambda_{123}^{2\nu} - \frac{1}{2} \Lambda_1^{2\nu} \Lambda_{123}^{4\nu} + \frac{1}{6}\Lambda_{123}^{6\nu} \ ,
\eea
and
\bea\label{ea:Lambdas}
\Lambda_{12} &=& \min[\Lambda_1,\Lambda_2] = \min(k_3^{-1}, |\vec p - \vec k_3|^{-1}, p^{-1},k_2^{-1}, |\vec p + \vec k_1|^{-1})\nn
\Lambda_{123} &=& \min[\Lambda_1,\Lambda_2,\Lambda_3] = \min(k_3^{-1}, |\vec p - \vec k_3|^{-1}, p^{-1},k_2^{-1}, |\vec p + \vec k_1|^{-1},k_1^{-1}) \ .
\eea

The $p$ integral is also controlled by the infrared contributions. One can see that the integrant of the $p$ integral has three branch points, and the integral is supported by the domain around these branch points. For example, for the branch point at $p = 0$, the integral is mainly supported at region around $p \lesssim \nu \times \min(k_1,k_3)$. Therefore, one can see that the three regions are well separated. Therefore, the leading order contribution of the integral can be separated to three parts that
\bea
&&\int \frac{d^3 p }{(2\pi)^3}  \frac{{\cal I}_3(\Lambda_1, \Lambda_2, \Lambda_3)}{|\vec p + \vec k_1|^{3 - 2 \nu}|\vec p - \vec k_3|^{3- 2 \nu}p^{3 - 2\nu}} \nn
&\simeq& \frac{1}{4\pi^2 \nu} \left\{ \frac{{\cal I}_3(\Lambda_1, \Lambda_2, \Lambda_3)|_{p\rightarrow0}}{ k_1^{3- 2\nu} k_3^{3 - 2\nu} \min(k_1,k_3)^{-2\nu}} + \frac{{\cal I}_3(\Lambda_1, \Lambda_2, \Lambda_3)|_{\vec p\rightarrow -\vec k_1}}{ k_1^{3- 2\nu} k_2^{3 - 2\nu} \min(k_1,k_2)^{-2\nu}}\right.\nn
&&\;\;\;\;\;\;\;\;\left.+\frac{{\cal I}_3(\Lambda_1, \Lambda_2, \Lambda_3)|_{\vec p\rightarrow \vec k_3}}{ k_2^{3- 2\nu} k_3^{3 - 2\nu} \min(k_2,k_3)^{-2\nu}} \right\} \nn
&=& \frac{1}{4\pi^2 \nu} \left\{ \frac{\frac{1}{2} k_3^{-2\nu} {k^{-4\nu}_{\max}} - \frac{1}{3} {k^{-6\nu}_{\max}}}{ k_1^{3- 2\nu} k_3^{3 - 2\nu} \min(k_1,k_3)^{-2\nu}} + \frac{\frac{1}{6} k^{-6\nu}_{\max}}{ k_1^{3- 2\nu} k_2^{3 - 2\nu} \min(k_1,k_2)^{-2\nu}}\right.\nn
&&\;\;\;\;\;\;\;\;\left.+\frac{k_3^{-2\nu}\max(k_2,k_3)^{-2\nu} k^{-2\nu}_{\max} - \frac{1}{2} \max(k_2,k_3)^{-4\nu} k^{-2\nu}_{\max} - \frac{1}{2} k_3^{-2\nu} k^{-4\nu}_{\max} + \frac{1}{6} k_{\max}^{-6\nu} }{ k_2^{3- 2\nu} k_3^{3 - 2\nu} \min(k_2,k_3)^{-2\nu}} \right\} \nn \ ,
\eea
where $k_{\max}$ is short for $\max(k_1,k_2,k_3)$.

With the permutation (and noting that diagram $a$ dominates) after straightforward calculation one can show that the leading contribution to the bi-spectrum is
\bea\label{eq:B3general2}
{\cal B}_\zeta(\vec k_1 , \vec k_2 , \vec k_3) &=& \frac{\pi |a_0|^6 H^6}{128\Lambda^6} \frac{1}{(2\nu)^4} \nn
&&\!\!\!\!\!\!\!\!\!\!\!\!\!\!\!\!\!\!\!\!\!\!\!\!\!\!\times\left\{ \frac{\max(k_1,k_3)^{-2\nu} k_{\max}^{-2\nu}}{k_1^{3-2\nu} k_3^{3-2\nu}} + \frac{\max(k_2,k_3)^{-2\nu} k_{\max}^{-2\nu}}{k_2^{3-2\nu} k_3^{3-2\nu}} + \frac{\max(k_1,k_2)^{-2\nu} k_{\max}^{-2\nu}}{k_1^{3-2\nu} k_2^{3-2\nu}}\right\}
\eea
Eq.~(\ref{eq:B3general2}) gives the general leading order expression for the bi-spectrum of $\zeta$ in the region that $m/H \ll1$.

\section{The tri-spectrum}
\label{sec:tri-spectrum}

There are in general three Feynman diagrams contributing to the one-loop four-point function of $\zeta$, which are shown in Fig.~\ref{fig:4pt}. Following Eq.~(\ref{eq:master}) the contribution from Fig.~\ref{fig:4pt}(a) can be written as
\bea
&&{\cal A}_4^{(a)} (\vec x_1, \vec x_2, \vec x_3, \vec x_4) \equiv \langle \zeta(\vec x_1, 0) \zeta(\vec x_1,0) \zeta(\vec x_3,0) \zeta(\vec x_4,0)\rangle^{(a)}  \nn
&=& \frac{1}{16\Lambda^8 H^8} \int_{-\infty}^{0} \frac{d \tau_4}{\tau_4^3} \int_{-\infty}^{\tau_4} \frac{d \tau_3}{\tau_3^3} \int_{-\infty}^{\tau_3} \frac{d \tau_2}{\tau_2^3} \int_{-\infty}^{\tau_2} \frac{d \tau_1}{\tau_1^3} \int d^3 y_1d^3 y_2d^3 y_3 d^3 y_4\nn
&&\times \langle \left[\pi'(\tau_1,\vec y_1)s^2(\tau_1,\vec y_1),\left[\pi'(\tau_2,\vec y_2)s^2(\tau_2,\vec y_2),\left[\pi'(\tau_3,\vec y_3)s^2(\tau_3,\vec y_3), \left[\pi'(\tau_4,\vec y_4)s^2(\tau_4,\vec y_4), \right.\right.\right.\right.\nn
&&\;\;\;\;\;\;\;   \left.\left.\left.\left.\pi(0,\vec x_1)\pi(0,\vec x_2)\pi(0,\vec x_3) \pi(0,\vec x_4)\right]\right]\right]\right] \rangle 
\eea

Let's follow the similar procedure calculating the integrant from inside out. 
\bea
{\cal S}_1&\equiv&\left[ \pi'(\tau_4,\vec y_4)s^2(\tau_4,\vec y_4), \pi(0,\vec x_1)\pi(0,\vec x_2)\pi(0,\vec x_3) \pi(0,\vec x_4) \right] \nn
&=& 2i {\rm Im}\langle \pi'(\tau_4,\vec y_4) \pi(0,\vec x_1)\rangle s^2(\tau_4,\vec y_4) \pi(0,\vec x_1)\pi(0,\vec x_2)\pi(0,\vec x_3) \nn
&&+ (x_1\leftrightarrow x_2) + (x_1\leftrightarrow x_3) + (x_1\leftrightarrow x_4) \ .
\eea
From Fig.~\ref{fig:4pt}(a) we can see that there are two options when we calculate 
\beq
{\cal S}_2 \equiv \left[ \pi'(\tau_3,\vec y_3)s^2(\tau_3,\vec y_3), {\cal S}_1 \right] \ 
\eeq
that the position $(\tau_3,\vec y_3)$ can be either next to or in diagonal to the position $(\tau_4,\vec y_4)$. Therefore, for the contribution from Fig.~\ref{fig:4pt}(a) we have 
\bea 
{\cal S}_2^{(1)} &=& 8i {\rm Im}\langle \pi'(\tau_4,\vec y_4) \pi(0,\vec x_1)\rangle\nn
&& \times \left( \langle \left[\pi'(\tau_3,\vec y_3)s(\tau_3,\vec y_3),\pi(0,\vec x_2)s(\tau_4,\vec y_4)\right]\rangle s(\tau_3,\vec y_3) s(\tau_4,\vec y_4) \pi(0,\vec x_3) \pi(0, \vec x_4)\right. \nn
&& +\left. (x_2\leftrightarrow x_3) + (x_2 \leftrightarrow x_4)\right)\nn
&&+(x_1 \leftrightarrow x_2) + (x_1 \leftrightarrow x_3) + (x_1 \leftrightarrow x_4) \nn
&=& -16 \left[ {\rm Im}\langle \pi'(\tau_4,\vec y_4) \pi(0,\vec x_1)\rangle {\rm Im}\left(\langle \pi'(\tau_3,\vec y_3) \pi(0, \vec x_2)\rangle \langle s(\tau_3,\vec y_3) s(\tau_4,\vec y_4)\rangle  \right)\right. \nn
&&\times s(\tau_3,\vec y_3) s(\tau_4,\vec y_4) \pi(0,\vec x_3) \pi(0, \vec x_4)
 +\left. (x_2\leftrightarrow x_3) + (x_2 \leftrightarrow x_4)\right]\nn
&&+(x_1 \leftrightarrow x_2) + (x_1 \leftrightarrow x_3) + (x_1 \leftrightarrow x_4) \ .
\eea
\bea
{\cal S}_2^{(2)} &=& - 4 
\left[ {\rm Im}\langle \pi'(\tau_4,\vec y_4) \pi(0,\vec x_1)\rangle  {\rm Im}\langle \pi'(\tau_3,\vec y_3) \pi(0,\vec x_2)\rangle s^2(\tau_3,\vec y_3) s^2(\tau_4,\vec y_4) \pi(0,\vec x_3) \pi(0, \vec x_4)\right.\nn
&&+\left. (x_2\leftrightarrow x_3) + (x_2 \leftrightarrow x_4)\right] +(x_1 \leftrightarrow x_2) + (x_1 \leftrightarrow x_3) + (x_1 \leftrightarrow x_4) \ .
\eea

Now let's calculate 
\beq
{\cal S}_3 \equiv [\pi'(\tau_2,\vec y_2) s^2(\tau_2,\vec y_2), {\cal S}_2] \ .
\eeq
The calculation is straightforward, we have
\bea
{\cal S}_3^{(1)} &=& -64i {\rm Im}\langle \pi'(\tau_4,\vec y_4) \pi(0,\vec x_1)\rangle {\rm Im}\left(\langle \pi'(\tau_3,\vec y_3) \pi(0, \vec x_2)\rangle \langle s(\tau_3,\vec y_3) s(\tau_4,\vec y_4)\rangle  \right)\nn
&&\times \left[ {\rm Im}(\langle s(\tau_2,\vec y_2) s(\tau_3,\vec y_3)\rangle \langle \pi'(\tau_2,\vec y_2)\pi(0,\vec x_3)\rangle) s(\tau_2,\vec y_2) s(\tau_4,\vec y_4) \pi(0, \vec x_4) \right. \nn
&& +  \left.{\rm Im}(\langle s(\tau_2,\vec y_2) s(\tau_4,\vec y_4)\rangle \langle \pi'(\tau_2,\vec y_2)\pi(0,\vec x_3)\rangle) s(\tau_2,\vec y_2) s(\tau_3,\vec y_3) \pi(0, \vec x_4) \right] \nn
&& + {\rm per}(x_1, x_2, x_3, x_4) \ .
\eea
\bea
{\cal S}_3^{(2)} &=& -64i {\rm Im}\langle \pi'(\tau_4,\vec y_4) \pi(0,\vec x_1)\rangle  {\rm Im}\langle \pi'(\tau_3,\vec y_3) \pi(0,\vec x_2)\rangle\nn
&&\times  {\rm Im}\left( \langle s(\tau_2,\vec y_2) s(\tau_3,\vec y_3)\rangle \langle s(\tau_2,\vec y_2) s(\tau_4, \vec y_4)\rangle \langle \pi'(\tau_2, \vec y_2) \pi(0,\vec x_3) \right) s(\tau_3,\vec y_3) s(\tau_4,\vec y_4) \pi(0,\vec x_4) \nn
&&+ {\rm per}(x_1, x_2, x_3, x_4) \ .
\eea
Therefore we have
\bea
{\cal S}_4 &\equiv& [\pi'(\tau_1,\vec y_1) s^2(\tau_1,\vec y_1) , {\cal S}_3] \nn
&=& 256 \times \left[ {\rm Im}\langle \pi'(\tau_4,\vec y_4) \pi(0,\vec x_1)\rangle {\rm Im}\left(\langle \pi'(\tau_3,\vec y_3) \pi(0, \vec x_2)\rangle \langle s(\tau_3,\vec y_3) s(\tau_4,\vec y_4)\rangle  \right) \right.\nn
&&\;\;\;\;\;\;\times {\rm Im}(\langle s(\tau_2,\vec y_2) s(\tau_3,\vec y_3)\rangle \langle \pi'(\tau_2,\vec y_2)\pi(0,\vec x_3)\rangle) \nn
&&\;\;\;\;\;\; \times {\rm Im}\left( \langle s(\tau_1,\vec y_1) s(\tau_2,\vec y_2)\rangle \langle s(\tau_1,\vec y_1) s(\tau_4, \vec y_4)\rangle \langle \pi'(\tau_1,\vec y_1) \pi(0,\vec x_4) \right) \nn
&&\;\;\;\;\;+ {\rm Im}\langle \pi'(\tau_4,\vec y_4) \pi(0,\vec x_1)\rangle {\rm Im}\left(\langle \pi'(\tau_3,\vec y_3) \pi(0, \vec x_2)\rangle \langle s(\tau_3,\vec y_3) s(\tau_4,\vec y_4)\rangle  \right) \nn
&&\;\;\;\;\;\;\times {\rm Im}(\langle s(\tau_2,\vec y_2) s(\tau_4,\vec y_4)\rangle \langle \pi'(\tau_2,\vec y_2)\pi(0,\vec x_3)\rangle) \nn
&&\;\;\;\;\;\; \times {\rm Im}\left( \langle s(\tau_1,\vec y_1) s(\tau_2,\vec y_2)\rangle \langle s(\tau_1,\vec y_1) s(\tau_3, \vec y_3)\rangle \langle \pi'(\tau_1,\vec y_1) \pi(0,\vec x_4) \right) \nn
&&\;\;\;\;\;+ {\rm Im}\langle \pi'(\tau_4,\vec y_4) \pi(0,\vec x_1)\rangle  {\rm Im}\langle \pi'(\tau_3,\vec y_3) \pi(0,\vec x_2)\rangle\nn
&&\;\;\;\;\;\; \times  {\rm Im}\left( \langle s(\tau_2,\vec y_2) s(\tau_3,\vec y_3)\rangle \langle s(\tau_2,\vec y_2) s(\tau_4, \vec y_4)\rangle \langle \pi'(\tau_2, \vec y_2) \pi(0,\vec x_3) \right) \nn
&&\;\;\;\;\;\; \times {\rm Im}\left( \langle s(\tau_1,\vec y_1)s(\tau_3,\vec y_3)\rangle \langle s(\tau_1,\vec y_1) s(\tau_4, \vec y_4)\rangle \langle \pi'(\tau_1,\vec y_1) \pi(0, \vec x_4) \right) \nn
&&+ {\rm per} (x_1,x_2, x_3, x_4)
\eea

Therefore we have 
\bea
&&{\cal A}_4^{(a)} (\vec x_1, \vec x_2, \vec x_3, \vec x_4) \nn
&=& \frac{16}{\Lambda^8 H^8} \int_{-\infty}^{0} \frac{d \tau_4}{\tau_4^3} \cdots \int_{-\infty}^{\tau_2} \frac{d \tau_1}{\tau_1^3} \int\frac{d^3k_1}{(2\pi)^3} \cdots \int\frac{d^3k_4}{(2\pi)^3}(2\pi)^4 \delta^4(\vec k_1 + \vec k_2 + \vec k_3 + \vec k_4)  e^{i (\vec k_1 \cdot \vec x_1 + \cdots + \vec k_4 \cdot \vec x_4)} \nn
&&\times\int\frac{d^3p}{(2\pi)^3} \left\{ {\rm Im} [\pi'_{k_1}(\tau_4) \pi^*_{k_1}(0)] {\rm Im}[\pi'_{k_2}(\tau_3) \pi^*_{k_2}(\tau_3) s_{|\vec p - \vec k_1|}(\tau_3) s_{|\vec p - \vec k_1|}^*(\tau_4) ] \right. \nn  
&&{\rm Im}[ \pi'_{k_3}(\tau_2) \pi^*_{k_3}(0)s_{|\vec p - \vec k_1 - \vec k_2|}(\tau_2) s_{|\vec p - \vec k_1 - \vec k_2|}^*(\tau_3) ]  {\rm Im}[\pi_{k_4}(\tau_1) \pi_{k_4}^*(0) s_{|\vec p + \vec k_4|}(\tau_1) s^*_{|\vec p + \vec k_4|)}(\tau_2) s_p(\tau_1) s_p^*(\tau_4)] \nn
&&+{\rm Im} [\pi'_{k_1}(\tau_4) \pi^*_{k_1}(0)] {\rm Im}[\pi'_{k_2}(\tau_3) \pi^*_{k_2}(\tau_3) s_{|\vec p - \vec k_2|}(\tau_3) s_{|\vec p - \vec k_2|}^*(\tau_4) ]  \nn  
&&{\rm Im}[ \pi'_{k_3}(\tau_2) \pi^*_{k_3}(0)s_{|\vec p - \vec k_1 - \vec k_2|}(\tau_2) s_{|\vec p - \vec k_1 - \vec k_2|}^*(\tau_4) ]  {\rm Im}[\pi_{k_4}(\tau_1) \pi_{k_4}^*(0) s_{|\vec p + \vec k_4|}(\tau_1) s^*_{|\vec p + \vec k_4|)}(\tau_2) s_p(\tau_1) s_p^*(\tau_3)] \nn
&&+{\rm Im}[\pi'_{k_1}(\tau_4)\pi^*_{k_1}(0) ] {\rm Im}[\pi'_{k_3}(\tau_2) \pi^*_{k_3}(0) s_{|\vec p + \vec k_2 + \vec k_4|}(\tau_2) s^*_{|\vec p + \vec k_2 + \vec k_4|}(\tau_3) s_{|\vec p - \vec k_1|}(\tau_2) s^*_{|\vec p - \vec k_1|}(\tau_4) ] \nn
&&\left.{\rm Im}[\pi'_{k_2}(\tau_3)\pi_{k_1}^*(0)] {\rm Im}[\pi'_{k_4}(\tau_1)\pi^*_{k_4}(0) s_{|\vec p + \vec k_4|}(\tau_1) s_{|\vec p + \vec k_4|}^*(\tau_3) s_p(\tau_1) s_p^*(\tau_4)] \right\}\nn
&& + {\rm per}(x_1, x_2, x_3, x_4) \ .
\eea
The tri-spectrum of $\zeta$ can be defined as
\beq
{\cal A}_4(\vec x_1, \vec x_2, \vec x_3, \vec x_4) = \int \frac{d^3k_1}{(2\pi)^3} \cdots\frac{d^3k_4}{(2\pi)^3} e^{i(\vec k_1 \cdot \vec x_1 + \cdots \vec k_4 \cdot \vec x_4)} (2\pi)^3 \delta^3(\vec k_1 + \cdots + \vec k_4) {\cal B}_{\zeta} (\vec k_1, \vec k_2, \vec k_3, \vec k_4) \ .
\eeq
Therefore noting that diagram ($a$) dominates, 
\bea\label{eq:B4pt}
&&{\cal B}_{\zeta} (\vec k_1, \vec k_2, \vec k_3, \vec k_4) \nn
&=& \frac{16}{\Lambda^8 H^8} \int_{-\infty}^{0} \frac{d \tau_4}{\tau_4^3} \cdots \int_{-\infty}^{\tau_2} \frac{d \tau_1}{\tau_1^3} \frac{d^3p}{(2\pi)^3} 
\nn
&&\left\{ {\rm Im} [\pi'_{k_1}(\tau_4) \pi^*_{k_1}(0)] {\rm Im}[\pi'_{k_2}(\tau_3) \pi^*_{k_2}(0) s_{|\vec p - \vec k_1|}(\tau_3) s_{|\vec p - \vec k_1|}^*(\tau_4) ] \right. \nn  
&&{\rm Im}[ \pi'_{k_3}(\tau_2) \pi^*_{k_3}(0)s_{|\vec p - \vec k_1 - \vec k_2|}(\tau_2) s_{|\vec p - \vec k_1 - \vec k_2|}^*(\tau_3) ]  {\rm Im}[\pi_{k_4}(\tau_1) \pi_{k_4}^*(0) s_{|\vec p + \vec k_4|}(\tau_1) s^*_{|\vec p + \vec k_4|)}(\tau_2) s_p(\tau_1) s_p^*(\tau_4)] \nn
&&+{\rm Im} [\pi'_{k_1}(\tau_4) \pi^*_{k_1}(0)] {\rm Im}[\pi'_{k_2}(\tau_3) \pi^*_{k_2}(\tau_3) s_{|\vec p - \vec k_2|}(\tau_3) s_{|\vec p - \vec k_2|}^*(\tau_4) ]  \nn  
&&{\rm Im}[ \pi'_{k_3}(\tau_2) \pi^*_{k_3}(0)s_{|\vec p - \vec k_1 - \vec k_2|}(\tau_2) s_{|\vec p - \vec k_1 - \vec k_2|}^*(\tau_4) ]  {\rm Im}[\pi_{k_4}(\tau_1) \pi_{k_4}^*(0) s_{|\vec p + \vec k_4|}(\tau_1) s^*_{|\vec p + \vec k_4|)}(\tau_2) s_p(\tau_1) s_p^*(\tau_3)] \nn
&&+{\rm Im}[\pi'_{k_1}(\tau_4)\pi^*_{k_1}(0) ] {\rm Im}[\pi'_{k_3}(\tau_2) \pi^*_{k_3}(0) s_{|\vec p + \vec k_2 + \vec k_4|}(\tau_2) s^*_{|\vec p + \vec k_2 + \vec k_4|}(\tau_3) s_{|\vec p - \vec k_1|}(\tau_2) s^*_{|\vec p - \vec k_1|}(\tau_4) ] \nn
&&\left.{\rm Im}[\pi'_{k_2}(\tau_3)\pi_{k_1}^*(0)] {\rm Im}[\pi'_{k_4}(\tau_1)\pi^*_{k_4}(0) s_{|\vec p + \vec k_4|}(\tau_1) s_{|\vec p + \vec k_4|}^*(\tau_3) s_p(\tau_1) s_p^*(\tau_4)] \right\}\nn
&& + {\rm per}(k_1, k_2, k_3, k_4) \ .
\eea
Just like in the case of the three-point correlation function of $\zeta$, the all the $\tau_i$ integrals are mainly supported in the infrared region. Therefore, to get the leading contribution one can again Laurent expand the mode functions of $\pi$ and $s$ and collect the leading terms. Follow the notations in Eq.~(\ref{eq:H2}) we have to the leading order
\bea
{\cal B}_{\zeta} &\simeq& \frac{\pi^4 H^8 |a_0|^8}{256 \Lambda^8} \int\frac{d^3p}{(2\pi)^3} \left\{ \frac{1}{p^{3 - 2\nu} |\vec p - \vec k_1|^{3-2\nu} |\vec p - \vec k_1 - \vec k_2|^{3-2\nu} |\vec p + \vec k_4|^{3-2\nu} }\right.\nn
&&\times \int_{-\Lambda_4^{(1)}}^0 \frac{d\tau_4}{(-\tau_4)^{1- 2\nu}} \int_{-\Lambda_3^{(1)}}^{\tau_4} \frac{d\tau_3}{(-\tau_3)^{1- 2\nu}} \int_{-\Lambda_2^{(1)}}^{\tau_3} \frac{d\tau_2}{(-\tau_2)^{1- 2\nu}} \int_{-\Lambda_1^{(1)}}^{\tau_2} \frac{d\tau_1}{(-\tau_1)^{1- 2\nu}}  \nn
&&+ \frac{1}{p^{3 - 2\nu} |\vec p - \vec k_2|^{3-2\nu} |\vec p - \vec k_1 - \vec k_2|^{3-2\nu} |\vec p + \vec k_4|^{3-2\nu} }\nn
&&\times \int_{-\Lambda_4^{(2)}}^0 \frac{d\tau_4}{(-\tau_4)^{1- 2\nu}} \int_{-\Lambda_3^{(2)}}^{\tau_4} \frac{d\tau_3}{(-\tau_3)^{1- 2\nu}} \int_{-\Lambda_2^{(2)}}^{\tau_3} \frac{d\tau_2}{(-\tau_2)^{1- 2\nu}} \int_{-\Lambda_1^{(2)}}^{\tau_2} \frac{d\tau_1}{(-\tau_1)^{1- 2\nu}}  \nn
&&+ \frac{1}{p^{3 - 2\nu} |\vec p - \vec k_1|^{3-2\nu} |\vec p + \vec k_2 + \vec k_4|^{3-2\nu} |\vec p + \vec k_4|^{3-2\nu} }\nn
&&\left.\times \int_{-\Lambda_4^{(3)}}^0 \frac{d\tau_4}{(-\tau_4)^{1- 2\nu}} \int_{-\Lambda_3^{(3)}}^{\tau_4} \frac{d\tau_3}{(-\tau_3)^{1- 2\nu}} \int_{-\Lambda_2^{(3)}}^{\tau_3} \frac{d\tau_2}{(-\tau_2)^{1- 2\nu}} \int_{-\Lambda_1^{(3)}}^{\tau_2} \frac{d\tau_1}{(-\tau_1)^{1- 2\nu}} \right\} \nn
&& + {\rm per} (k_1, k_2, k_3, k_4 ) \ , \nn
&=& \frac{\pi^4 H^8 |a_0|^8}{256 \Lambda^8} \left(\frac{1}{2\nu}\right)^4 \int\frac{d^3p}{(2\pi)^3} \left\{  \frac{{\cal I}(\Lambda_1^{(1)},\Lambda_2^{(1)},\Lambda_3^{(1)},\Lambda_4^{(1)})}{p^{3 - 2\nu} |\vec p - \vec k_1|^{3-2\nu} |\vec p - \vec k_1 - \vec k_2|^{3-2\nu} |\vec p + \vec k_4|^{3-2\nu} } \right. \nn
&& \;\;\;\;\;\;\;\;\;\;\;\;+ \frac{{\cal I}(\Lambda_1^{(2)},\Lambda_2^{(2)},\Lambda_3^{(2)},\Lambda_4^{(2)})}{p^{3 - 2\nu} |\vec p - \vec k_2|^{3-2\nu} |\vec p - \vec k_1 - \vec k_2|^{3-2\nu} |\vec p + \vec k_4|^{3-2\nu} }\nn
&&  \;\;\;\;\;\;\;\;\;\;\;\;\left.+\frac{{\cal I}(\Lambda_1^{(3)},\Lambda_2^{(3)},\Lambda_3^{(3)},\Lambda_4^{(3)})}{p^{3 - 2\nu} |\vec p - \vec k_1|^{3-2\nu} |\vec p + \vec k_2 + \vec k_4|^{3-2\nu} |\vec p + \vec k_4|^{3-2\nu} }\right\} + {\rm per}(k_1, k_2, k_3, k_4)\ .
\eea
The UV cutoffs $\Lambda_i^{(j)}$ can be read out directly from Eq.~(\ref{eq:B4pt}) that
\bea
&&\Lambda_1^{(1)} = \min(k_4^{-1},p^{-1}, |\vec p + \vec k_4|^{-1}),\; \Lambda_2^{(1)} = \min(k_3^{-1},|\vec p - \vec k_1 - \vec k_2|^{-1}, |\vec p + \vec k_4|^{-1}) \nn
&&\Lambda_3^{(1)} = \min(k_2^{-1},|\vec p - \vec k_1|^{-1}, |\vec p - \vec k_1- \vec k_2|),\; \Lambda_4^{(1)} = \min(k_1, |\vec p - \vec k_1|^{-1}, p^{-1}) \ ; \nn
&&\Lambda_1^{(2)} = \min(k_4^{-1},p^{-1}, |\vec p + \vec k_4|^{-1}),\; \Lambda_2^{(2)} = \min(k_3^{-1},|\vec p - \vec k_1 - \vec k_2|^{-1}, |\vec p + \vec k_4|^{-1}) \nn
&&\Lambda_3^{(2)} = \min(k_2^{-1},|\vec p - \vec k_2|^{-1}, p^{-1}),\; \Lambda_4^{(2)} = \min(k_1, |\vec p - \vec k_2|^{-1}, |\vec p - \vec k_1 - \vec k_2|^{-1}) \ ; \nn
&&\Lambda_1^{(3)} = \min(k_4^{-1},p^{-1}, |\vec p + \vec k_4|^{-1}),\; \Lambda_2^{(3)} = \min(k_3^{-1},|\vec p + \vec k_2 + \vec k_4|^{-1}, |\vec p - \vec k_1|^{-1}) \nn
&&\Lambda_3^{(3)} = \min(k_2^{-1},|\vec p + \vec k_4|^{-1}, |\vec p + \vec k_2 + \vec k_4|),\; \Lambda_4^{(3)} = \min(k_1, |\vec p - \vec k_1|, p^{-1}) \ ,
\eea
and 
\bea
{\cal I} (\Lambda_1, \Lambda_2, \Lambda_3, \Lambda_4)  &=& -\frac{1}{2} \Lambda _{123}^{4 \nu } \Lambda _{1234}^{2 \nu } \Lambda _1^{2 \nu }+\frac{1}{6} \Lambda _{1234}^{6 \nu } \Lambda _1^{2 \nu }-\frac{1}{2} \Lambda _{12}^{4 \nu } \Lambda _{123}^{2 \nu } \Lambda _{1234}^{2 \nu }+\left(\Lambda _1 \Lambda _{12} \Lambda _{123} \Lambda _{1234}\right){}^{2 \nu } \nn
&&+\frac{1}{6} \Lambda _{123}^{6 \nu } \Lambda _{1234}^{2 \nu }-\frac{1}{2}  \left(\Lambda _1^{2 \nu } \Lambda _{12}^{2 \nu } \Lambda _{1234}^{4 \nu }\right)+\frac{1}{4} \Lambda _{12}^{4 \nu } \Lambda _{1234}^{4 \nu }-\frac{1}{24}\Lambda _{1234}^{8 \nu } \ .
\eea
The definition of $\Lambda^{(i)}_{12}$, $\Lambda^{(i)}_{123}$ and $\Lambda^{(i)}_{1234}$ are like in Eq.~(\ref{ea:Lambdas}). Similar to calculation of the tri-spectrum, one can capture the dominant contribution of the $p$-integral. This gives Eq.~(\ref{eq:3_4}) as the leading order result for the tri-spectrum with a general wave vector configuration.  
%We can get
%\bea
%&&{\cal B}_{\zeta}(\vec k_1, \vec k_2, \vec k_3, \vec k_4) \nn
%&=& \frac{\pi^2 H^8 |a_0|^8}{1024 \Lambda^8} \left(\frac{1}{2\nu}\right)^5\nn
%&&\times 
%\frac{\min[\max(k_1,k_2,|\vec k_1 + \vec k_1|),\max(k_3, k_4,|\vec k_3 + \vec k_4|)]^{-2\nu} \max(k_1,k_2,k_3,k_4,|\vec k_1 + \vec k_2|)^{-2\nu}}{k_1^3 k_4^3 |\vec k_1 + \vec k_2|^{3-2\nu}  \min(k_1,k_4, |\vec k_1 + \vec k_2|)^{-2\nu}} \nn
%&& + {\rm per}(k_1, k_2, k_3, k_4) \ .
%\eea
%This gives the leading order result for the tri-spectrum with a general wave vector configuration. 

\section{One-loop correction to two-point function of $\zeta$}
\label{app:d}

\begin{figure}
\centering
\includegraphics[height=1.in]{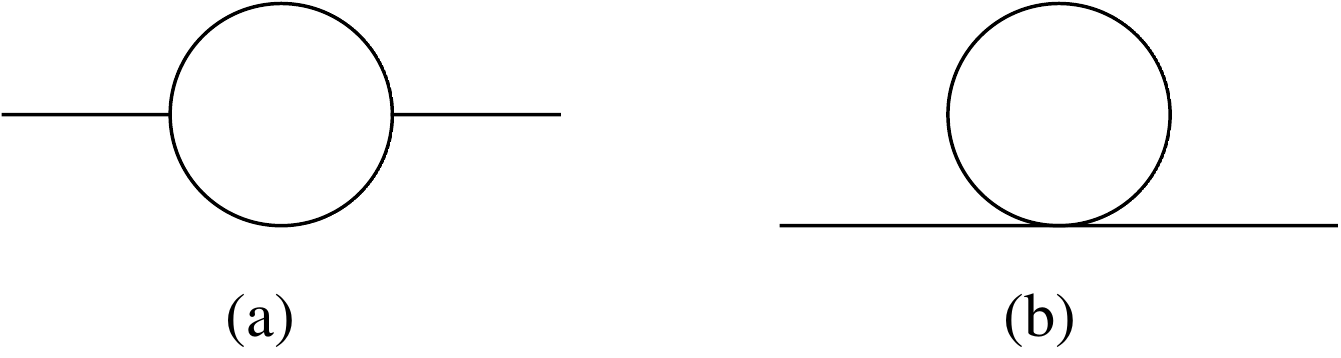}
\caption{}\label{fig:2pt}
\end{figure}

The Feynman diagrams contributing to the one-loop correction to the two-point function of $\zeta$ are shown in Fig.~\ref{fig:2pt}. Similar to the cases of the three and four-point functions the contribution from Fig.~\ref{fig:2pt}(a) dominates. Its contribution can be written as
\bea
{\cal A}_2^{(a)}(\vec x_1 - \vec x_2) &=& - \frac{1}{4 H^4 \dot \Lambda^4} \int_{-\infty}^0 \frac{d\tau_2}{\tau_2^3} \int_{-\infty}^{\tau_2} \frac{d\tau_1}{\tau_1^3} \int d^3 y_1 \int d^3 y_2 \nn 
&& \times \langle [\pi'(\tau_1,\vec y_1) s^2(\tau_1,\vec y_1), [\pi'(\tau_2,\vec y_2) s^2(\tau_2,\vec y_2),\pi(0,\vec x_1)\pi(0,\vec x_2)]] \rangle \nn
&=& - \frac{1}{4 H^4 \dot \Lambda^4} \int_{-\infty}^0 \frac{d\tau_2}{\tau_2^3} \int_{-\infty}^{\tau_2} \frac{d\tau_1}{\tau_1^3} \int d^3 y_1 \int d^3 y_2 \nn
&&\times \left( \langle [\pi'(\tau_1,\vec y_1) s^2(\tau_1,\vec y_1),s^2(\tau_2,\vec y_2) \pi(0,\vec x_2)]\rangle \langle [\pi'(\tau_2,\vec y_2),\pi(0,\vec x_1)] + (x_1 \leftrightarrow x_2)\right) \nn
&=& - \frac{1}{4 H^4 \dot \Lambda^4} \int_{-\infty}^0 \frac{d\tau_2}{\tau_2^3} \int_{-\infty}^{\tau_2} \frac{d\tau_1}{\tau_1^3} \int d^3 y_1 \int d^3 y_2 \nn
&&\times (-8) \left({\rm Im}\langle\pi'(\tau_2,\vec y_2) \pi(0,\vec x_2)\rangle {\rm Im}\left[\langle \pi'(\tau_1,\vec y_1) \pi(0,\vec x_1) \rangle \langle s(\tau_1,\vec y_1) s(\tau_2,\vec y_2)\rangle^2 \right]+(x_1\leftrightarrow x_2)\right) \nn
&\approx& \frac{\pi^2 H^4 |a_0|^4}{16\Lambda^4} \int\frac{d^3k}{(2\pi)^3} \frac{1}{k^2} \cos[\vec k\cdot (\vec x_1 - \vec x_2)] \int_{-\infty}^0 \frac{d\tau_2}{\tau_2^3} \int_{-\infty}^{\tau_2} \frac{d\tau_1}{\tau_1^3} \frac{d^3p}{(2\pi^3)} \nn
&&\times (-\tau_1)^{1+2\nu} (-\tau_2)^{1+2\nu} \sin(k\tau_1) \sin(k\tau_2) \frac{1}{p^{3-2\nu} |\vec k - \vec p|^{3-2\nu}} \nn
&=& \frac{H^4 |a_0|^4}{256\nu^3 \Lambda^4} \int \frac{d^3k}{(2\pi)^3} \frac{1}{k^3} e^{i \vec k \cdot (\vec x_1 - \vec x_2) } \ .
\eea
From this one can read out the correction to the curvature perturbation 
\bea
\delta\Delta_\zeta^2 = \frac{H^4 |a_0|^4}{512 \pi^2 \nu^3 \Lambda^4} \ .
\eea


\begin{thebibliography}{99}

\bibitem{SKS}
A.~A.~ Starobinsky, JETP Lett. {\bf 30}, 682 (1979);
A.~Guth, Phys. Rev. D{\bf 23}, 347 (1981); A.~D.~Linde, Phys. Lett. B {\bf 108}, 389 (1982); {\bf 114}, 431 (1982); A.~Albrecht and P.~Steinhardt, Phys. Rev. Lett. {\bf 48}, 1220 (1982).

 \bibitem{Maldacena:2002vr} 
  J.~M.~Maldacena,
  %``Non-Gaussian features of primordial fluctuations in single field inflationary models,''
  JHEP {\bf 0305}, 013 (2003)
  %doi:10.1088/1126-6708/2003/05/013
  [astro-ph/0210603].    
    
    \bibitem{AGW}
  T.~J.~Allen, B.~Grinstein and M.~B.~Wise, Phys.\ Lett.\ B. {\bf 197}, 66 (1987).
   
 
\bibitem{Dalal:2007cu} 
  N.~Dalal, O.~Dor\'e, D.~Huterer and A.~Shirokov,
  %``The imprints of primordial non-gaussianities on large-scale structure: scale dependent bias and abundance of virialized objects,''
  Phys.\ Rev.\ D {\bf 77}, 123514 (2008)
  %doi:10.1103/PhysRevD.77.123514
  [arXiv:0710.4560 [astro-ph]].
  
  \bibitem{Baumann:2012bc} 
  D.~Baumann, S.~Ferraro, D.~Green and K.~M.~Smith,
  %``Stochastic Bias from Non-Gaussian Initial Conditions,''
  JCAP {\bf 1305}, 001 (2013)
  %doi:10.1088/1475-7516/2013/05/001
  [arXiv:1209.2173 [astro-ph.CO]].
  
  %\cite{Gleyzes:2016tdh}
\bibitem{Gleyzes:2016tdh} 
  J.~Gleyzes, R.~de Putter, D.~Green and O.~Dor\'e,
  %``Biasing and the search for primordial non-Gaussianity beyond the local type,''
  JCAP {\bf 1704}, no. 04, 002 (2017)
  %doi:10.1088/1475-7516/2017/04/002
  [arXiv:1612.06366 [astro-ph.CO]].
  

  %\cite{MoradinezhadDizgah:2017szk}
%\bibitem{MoradinezhadDizgah:2017szk} 
  %A.~Moradinezhad Dizgah and C.~Dvorkin,
  %``Scale-Dependent Galaxy Bias from Massive Particles with Spin during Inflation,''
  %arXiv:1708.06473 [astro-ph.CO].
    
    %\cite{An:2017rwo}
\bibitem{An:2017rwo} 
  H.~An, M.~McAneny, A.~K.~Ridgway and M.~B.~Wise,
  %``Non-Gaussian Enhancements of Galactic Halo Correlations in Quasi-Single Field Inflation,''
  arXiv:1711.02667 [hep-ph].
  %%CITATION = ARXIV:1711.02667;%%
  %7 citations counted in INSPIRE as of 25 May 2018
  
  \bibitem{Chen:2009zp} 
  X.~Chen and Y.~Wang,
  %``Quasi-Single Field Inflation and Non-Gaussianities,''
  JCAP {\bf 1004}, 027 (2010)
  %doi:10.1088/1475-7516/2010/04/027
  [arXiv:0911.3380 [hep-th]].
  %%CITATION = doi:10.1088/1475-7516/2010/04/027;%%
  %211 citations counted in INSPIRE as of 23 May 2017    

\bibitem{Goroff:1986ep}
  M.~H.~Goroff, B.~Grinstein, S.~J.~Rey and M.~B.~Wise,
  %``Coupling of Modes of Cosmological Mass Density Fluctuations,''
  Astrophys.\ J.\  {\bf 311}, 6 (1986).
   %\cite{Chen:2009zp}

    
%\cite{McAneny:2017bbv}
\bibitem{McAneny:2017bbv} 
  M.~McAneny, A.~K.~Ridgway, M.~P.~Solon and M.~B.~Wise,
  %``Loop-Induced Stochastic Bias at Small wave vectors,''
  arXiv:1712.07657 [astro-ph.CO].
  %%CITATION = ARXIV:1712.07657;%%
  
  \cite{Ade:2013ydc}
\bibitem{Ade:2013ydc} 
  P.~A.~R.~Ade {\it et al.} [Planck Collaboration],
  %``Planck 2013 Results. XXIV. Constraints on primordial non-Gaussianity,''
  Astron.\ Astrophys.\  {\bf 571}, A24 (2014)
   [arXiv:1303.5084 [astro-ph.CO]].
    
  \bibitem{Ade:2015ava} 
  P.~A.~R.~Ade {\it et al.} [Planck Collaboration],
  %``Planck 2015 results. XVII. Constraints on primordial non-Gaussianity,''
  Astron.\ Astrophys.\  {\bf 594}, A17 (2016)
  [arXiv:1502.01592 [astro-ph.CO]].
  %\cite{Gunn:1972sv}

\bibitem{Gunn:1972sv} 
  J.~E.~Gunn and J.~R.~Gott, III,
  %``On the Infall of Matter into Clusters of Galaxies and Some Effects on Their Evolution,''
  Astrophys.\ J.\  {\bf 176}, 1 (1972).
  doi:10.1086/151605
  %%CITATION = doi:10.1086/151605;%%
  %1737 citations counted in INSPIRE as of 25 May 2018
    
%\cite{Cheung:2007st}
\bibitem{Cheung:2007st} 
  C.~Cheung, P.~Creminelli, A.~L.~Fitzpatrick, J.~Kaplan and L.~Senatore,
  %``The Effective Field Theory of Inflation,''
  JHEP {\bf 0803}, 014 (2008)
  [arXiv:0709.0293 [hep-th]].
  %%CITATION = doi:10.1088/1126-6708/2008/03/014;%%
  %605 citations counted in INSPIRE as of 12 Jun 2018    
    
%\cite{Weinberg:2005vy}
\bibitem{Weinberg:2005vy} 
  S.~Weinberg,
  %``Quantum contributions to cosmological correlations,''
  Phys.\ Rev.\ D {\bf 72}, 043514 (2005)
  [hep-th/0506236].
  %%CITATION = doi:10.1103/PhysRevD.72.043514;%%
  %548 citations counted in INSPIRE as of 12 Jun 2018    
    
 \bibitem{Arkani-Hamed:2015bza} 
  N.~Arkani-Hamed and J.~Maldacena,
  %``Cosmological Collider Physics,''
  arXiv:1503.08043 [hep-th].
      
   
    
\end{thebibliography}
\end{document}